\documentclass[sigconf,authorversion]{acmart}

\usepackage{xspace}
\usepackage{booktabs}
\usepackage{graphicx}
\usepackage{pifont}
\usepackage{algorithm}
\usepackage{algorithmic}
\usepackage{enumitem}

\newcommand{\cmark}{\ding{51}}
\newcommand{\xmark}{\ding{55}}
\usepackage{fontawesome5}
\newcommand{\halfmark}{\faAdjust}
\usepackage{xcolor}
\newcommand{\unveil}{Unveil\xspace}
\newcommand{\gcmark}{\textcolor{green}{\cmark}}
\newcommand{\rxmark}{\textcolor{red}{\xmark}}

\AtBeginDocument{}

\copyrightyear{2026}
\acmYear{2026}
\setcctype{by}
\setcopyright{cc}
\acmBooktitle{ACM Symposium on Cloud Computing Volume 2 (SoCC '26 Vol. 2), November 18--20, 2026, Singapore, Singapore}
\acmDOI{10.1145/3828161.3857898}
\acmConference[SoCC '26 Vol. 2]{ACM Symposium on Cloud Computing Volume 2}{November 18--20, 2026}{Singapore, Singapore}
\acmISBN{979-8-4007-2816-7/2026/11}

\begin{document}

\title{Seeing is believing: The case for hardware-centric inefficiency detection and remediation}

\author{Ziji Chen}
\affiliation{\institution{University of Oxford} \country{United Kingdom}}
\email{ziji.chen@eng.ox.ac.uk}

\author{Steven W. D. Chien}
\affiliation{\institution{University of St Andrews} \country{United Kingdom}}
\email{steven.chien@st-andrews.ac.uk}

\author{Peng Qian}
\affiliation{\institution{University of Oxford} \country{United Kingdom}}
\email{peng.qian@eng.ox.ac.uk}

\author{Noa Zilberman}
\affiliation{\institution{University of Oxford} \country{United Kingdom}}
\email{noa.zilberman@eng.ox.ac.uk}

\ccsdesc[300]{Computer systems organization~Maintainability and maintenance}
\keywords{machine learning systems, cloud computing, silent inefficiency, anomaly detection, host-visible telemetry, fingerprinting}

\begin{abstract}
Large-scale machine learning workloads on the cloud suffer from \emph{silent inefficiencies}, which do not trigger failures or obvious anomalies, making them hard to detect and diagnose.
Existing diagnosis systems rely on guest-side instrumentation or dedicated healthy baselines.
These assumptions do not hold in modern production cloud environments, where operators cannot observe inside the guest, and a healthy baseline for each setup is impractical because of system diversity.
We present \textbf{Unveil}, a hardware-centric observability framework for detecting, identifying, and remediating silent inefficiencies using only host-visible telemetry.
\unveil requires neither guest instrumentation nor application modification.
It uses unsupervised learning models to detect inefficiencies and identifies known ones through a novel fingerprinting algorithm.
An extensible knowledge base containing curated remediation scripts is used to correct problems without interrupting workloads when possible.
We evaluate Unveil on distributed machine learning workloads, such as large language model serving, introducing less than 2\% monitoring overhead.
It successfully recovered from a 7\% throughput loss and reduced inference latency by up to 43.48\%.
These results demonstrate that host-visible telemetry is sufficient for practical and continuous diagnosis of production machine learning systems.
\end{abstract}

\maketitle

\section{Introduction}\label{sec:introduction}

Fail-slow is a behavior in which a component slows down the entire distributed system, without causing failure~\cite{limplock13,grayfailure17,onesizefitsnone24}.
Large-scale machine learning (ML) training and serving are especially prone to fail-slow.
DeepSeek reports that training DeepSeek-V3 consumed 2.788 million H800 GPU-hours at an estimated cost of \$5.576 million~\cite{deepseekai2025deepseekv3technicalreport}.
This means even a 5\% decrease in end-to-end runtime can save hundreds of thousands of dollars.

We focus on one class of fail-slow problems, which we refer to as \emph{silent inefficiencies}. \emph{Silent inefficiencies} often stem from system misconfigurations or degraded operating conditions, such as memory fragmentation~\cite{misconfig13}. Correcting these problems usually requires only addressing the underlying cause rather than replacing hardware or failing over to another machine. However, they are difficult to identify because they do not crash jobs, produce explicit errors, or trigger obvious utilization anomalies. Thus, workloads can appear healthy while silently degrading performance.

Modern ML workloads increasingly run inside virtual machines (VMs) or containers.
However, many existing diagnosis and mitigation techniques 
rely on cooperation from the guest software stack.
Some instrument the workload directly~\cite{jiang2024megascalescalinglargelanguage}.
Others instrument software stacks inside the guest, including customized collective communication libraries~\cite{aegis}, Compute Unified Device Architecture (CUDA) event hooks~\cite{jiang2024megascalescalinglargelanguage,xu2025eacgmnoninstrumentedperformancetracing}, and NVIDIA Collective Communications Library (NCCL)-level tracing~\cite{deng2025mycroft,xu2025eacgmnoninstrumentedperformancetracing}.
Even on dedicated hosts, cloud providers still expose a VM or container to the customer~\cite{eroica}.
As a result, operators can deploy these techniques only from the host, which prevents them from accessing the guest-level execution context needed and significantly reduces their effectiveness.

Existing approaches also rely heavily on dedicated baselines, which are equally difficult to obtain and unportable.
Establishing one requires running workloads to completion under known healthy conditions.
Modern ML jobs can occupy thousands of GPUs for weeks, making this process prohibitively expensive.
Cloud hardware is also heterogeneous and continuously refreshed.
As a result, healthy baselines must be rebuilt and validated repeatedly~\cite{ByteRobust2025}.
Meanwhile, machines in the same cluster are typically provisioned, imaged, and deployed together.
A configuration mistake can therefore affect an entire deployment batch from the start.
Consequently, there are neither healthy peers for cross-machine comparison nor healthy historical executions for temporal comparison.
Approaches that rely on labeled healthy samples~\cite{Prodigy10.1145/3581784.3607076} or historical execution traces~\cite{xia2025nodesentry} lose the reference they depend on.

Cloud providers attempt to detect \emph{silent inefficiencies} through offline validation.
For example, Alibaba Cloud checks hosts before handing them to a customer~\cite{aegis}.
However, these checks run only once before deployment, so issues that emerge later remain undetected.
Microsoft's SuperBench periodically evaluates production nodes using a comprehensive benchmark suite, but each run takes hours and requires taking the node out of service~\cite{aegis,superbench24}, making it too expensive for continuous use.

The scale of modern ML infrastructure makes manual diagnosis impractical. 
Alibaba Cloud reports that one large-model training deployment expanded by more than 40$\times$ within only 16 months~\cite{aegis}. 
Alibaba reports that a production deployment of 100,000 GPUs experienced 80 serious performance regressions requiring manual diagnosis over an 18-month period~\cite{eroica}. Diagnosing a single regression can take a team of engineers an entire month~\cite{eroica}.

Furthermore, existing diagnosis systems simplify identification, but their remediation strategy remains isolation~\cite{ByteRobust2025,aegis}.
This is suitable for hardware faults because replacing the affected machine restores performance.
However, the hardware is healthy in \emph{silent inefficiencies}, and performance can be restored by software changes.
Here, operators need more than an alert identifying a slow machine.

We present \textbf{\unveil}, a hardware-centric observability and remediation framework for modern cloud ML deployments.
\unveil's key insight is that \emph{silent inefficiencies} can be identified from host-visible hardware behavior without guest cooperation or a dedicated healthy baseline (\autoref{sec:backgroundmotivation}).
\unveil collects only host-visible telemetry using standard tools (\autoref{sec:problemspace}), and uses online unsupervised detection (\autoref{sec:Methodology}).
Once an anomaly is detected, \unveil attributes it to the most likely repairable cause via signature matching against a knowledge base of known inefficiencies and remediation scripts (\autoref{sec:bottleneck}).
 Once matched, \unveil recommends or automatically applies the corresponding remediation, and when possible, does so without interrupting running applications (\autoref{sec:auto-remediation}).
Unmatched anomalies are recorded for further investigation by engineers, who can validate the anomalies and develop new remediation scripts to expand the knowledge base.

In summary, we make the following contributions:
\begin{itemize}[leftmargin=*]
    \item We design and develop \textbf{\unveil}, a hardware-centric observability framework for identifying \emph{silent inefficiencies} using only host-visible telemetry (\autoref{sec:problemspace}).
    \item We design a diagnosis methodology that combines online anomaly detection with novel fingerprint extraction and matching. We further introduce a cross-platform \emph{knowledge base} that maps fingerprints to remediation actions (\autoref{sec:implementation}).
    \item We demonstrate that \unveil resolves \emph{silent inefficiencies} in real-world applications, avoiding up to 7\% throughput loss and reducing inference latency by up to 43.48\% (\autoref{sec:real-world}).
\end{itemize}

\section{Background and motivation}\label{sec:backgroundmotivation}

Despite fail-slow inefficiencies in distributed systems being widely studied over the last decade~\cite{limplock13,onesizefitsnone24,grayfailure17}, they remain challenging to diagnose in practice, which is further exacerbated by the scale and complexity of modern ML workloads.
We summarize recent work on ML system monitoring in~\autoref{tab:related_monitor_comparison}.

\begin{table*}[h]
\centering
\small
\caption{Comparison with existing monitoring and diagnosis tools.}
\vspace{-10pt}
\label{tab:related_monitor_comparison}
\begin{tabular}{lccccccccl}
\toprule
\textbf{Tool} & \textbf{VM-compat.} & \textbf{Cont. Deploy} & \textbf{No Code $\triangle$} & \textbf{HW Indep.} & \textbf{No-Base.} & \textbf{Scope} \\
\midrule
Plumber~\cite{kuchnik2022plumber} & \rxmark & \halfmark & \rxmark & \gcmark & \gcmark (A) & data loading \\
SuperBench~\cite{superbench24} & \gcmark & \rxmark & \gcmark & \gcmark & \rxmark (H) & component \\
Aegis~\cite{aegis} & \rxmark & \gcmark & \gcmark & \gcmark & \halfmark (P) & compute or communication \\
Greyhound~\cite{greyhound} & \rxmark & \gcmark & \gcmark & \gcmark & \halfmark (S+P) & compute or communication \\
Pulse~\cite{xiao2026pulse} & \gcmark & \gcmark & \gcmark & \rxmark & \halfmark (S+P) & compute or communication \\
Mycroft~\cite{deng2025mycroft} & \rxmark & \gcmark & \gcmark & \gcmark & \halfmark (P) & Process rank + RDMA or GPU (communication\ only) \\
Holmes~\cite{holmes} & \rxmark & \gcmark & \gcmark & \gcmark & \rxmark (H) & device (communication\ only) \\
eACGM~\cite{xu2025eacgmnoninstrumentedperformancetracing} & \rxmark & \halfmark & \gcmark & \gcmark & \gcmark (S) & metrics, layers \\
Prodigy~\cite{Prodigy10.1145/3581784.3607076} & \gcmark & \gcmark & \gcmark & \gcmark & \rxmark (H) & metrics (CPU/memory only) \\
\midrule
\textbf{\unveil} & \gcmark & \gcmark & \gcmark & \gcmark & \gcmark (S) & metrics, subsystems, \textbf{inefficiencies} \\
\bottomrule
\end{tabular}
\noindent\begin{minipage}{\textwidth}
\small{Note: VM-compat.\ = Host-only deployment for workloads in inaccessible VMs/containers; Cont. Deploy\ = Continuous Deployment; 
No Code $\triangle$\ = No User/Framework Code Changes; HW Indep.\ = No Special Hardware; No-Base.\ = No Dedicated Baseline Run for Anomaly Detection (H = pre-collected healthy baseline; P = runtime peer/cross-node/rank comparison; S = self-referential, using its own history; A = algorithmically derived expected value);}
\end{minipage}
\end{table*}

\subsection{Full stack complexity}
Modern ML workloads in a cloud environment are commonly deployed on virtualized instances or containers.
Within them, the actual application is developed and executed on a layered software stack that spans application frameworks, runtime systems, operating systems, and heterogeneous hardware.
Frameworks such as PyTorch~\cite{paszke2019pytorch} and TensorFlow~\cite{abadi2016tensorflow} abstract hardware details behind high-level execution graphs and operators, while vendor runtimes (e.g., CUDA) and drivers manage device scheduling, memory movement, and communication (e.g., NCCL).
At the hardware level, execution involves multi-socket CPUs, multiple accelerators per node, and high-bandwidth interconnects.
Observed symptoms at one layer may originate from inefficiencies elsewhere in the stack.
For example, a long GPU kernel launch time could stem from launching overhead rather than GPU saturation, and low GPU utilization is meaningless without correlating the application timeline~\cite{gleeson2021rl}.
Similarly, a slow input preprocessing pipeline could be a result of disk misconfiguration or inefficient caching, rather than a CPU processing bottleneck~\cite{kuchnik2022plumber}.
Fail-slow behavior is particularly problematic for ML workloads because these workloads progress synchronously, unlike traditional distributed systems.
This means that a slowdown in one component could instantly stall the entire ML system.

\subsection{Fine-grained profiling}
ML developers can use profilers provided by ML platforms, such as Torch profiler~\cite{pytorch-profiler} and TensorFlow profiler~\cite{tensorboard}, to perform fine-grained profiling.
These profilers have full access to the application and provide function-level tracing.
They also sample system-level metrics, such as CPU and GPU counters, to provide a full stack view, making them suitable for identifying software bugs and hotspots.
Unfortunately, they are unsuitable for detecting runtime system problems and misconfigurations in production because of high overhead.
For example, Torch Profiler~\cite{pytorch-profiler} incurs a 5\% runtime overhead and generates 0.7 GB of tracing data per hour.
Many problems, such as GPU thermal throttling, can happen intermittently during runtime, requiring constant monitoring.

Profilers fall short in virtualized or containerized environments, as what they observe on the guest side may not correspond to the host system.
In many cases, the guest simply cannot access those metrics and counters.
For example, hardware counters such as \texttt{cache misses} require both the \texttt{CAP\_PERFMON} capability and a permissive \texttt{perf\_event\_paranoid} kernel setting. Both require privilege and are restricted on shared nodes.
Eroica~\cite{eroica} overcomes this by sampling hardware counters from a privileged container on every host, then exporting them to other containers via a shared folder.
This raises security concerns and is impractical for operators in a production public cloud.
In fact, Aegis~\cite{aegis} does not even release the full list of metrics they used, citing confidentiality policies.

\subsection{Coarse-grained monitoring}
Alternatively, systems such as Greyhound~\cite{greyhound}, Aegis~\cite{aegis}, and Mycroft~\cite{deng2025mycroft} cross-reference metrics including iteration and collective communication operation time, which can be performed entirely by the guest.
However, this can only identify an offending node but not localize the problem. 
Meanwhile, a guest cannot directly migrate workloads or perform host-level placement changes.
As such, automatic remediation by Greyhound is limited to either changing training parameters or checkpoint-restart on another set of nodes.
Similarly, Aegis and SuperBench~\cite{superbench24} diagnose problems offline by benchmarking on the host.
Furthermore, systems that rely on collective communication operations primarily focus on network-related issues, but recent work finds that 71\% of distributed failures in production are irrelevant to the network~\cite{aegis}.

\subsection{Operator-level monitoring}
As a result, ML developers turn to the operators, who have full access to the host hardware.
Unfortunately, operator-level monitoring suffers from the opposite problem: operators have no visibility of the workload.
For example, eACGM~\cite{xu2025eacgmnoninstrumentedperformancetracing} monitors at the kernel with extended Berkeley Packet Filter (eBPF) to perform online anomaly detection.
Yet, it still requires hooking into userspace functions (e.g., \texttt{PyObject\_CallFunction}) and libraries (e.g., NCCL), making it infeasible for deployment outside a virtualized instance.
Systems such as Holmes~\cite {holmes} use communication operations to infer iteration time and delay in compute operations.
Pulse~\cite{xiao2026pulse} goes a step further and infers the type of parallelism by monitoring network traffic at a data processing unit.
However, it still requires overloading NCCL to monitor the start and end points of those operations on each process.
While these methods are non-intrusive to the user code, they require changes in libraries and focus on either compute or network. 
This is unreasonable in a cloud environment as it would require a user to deploy a particular image provided by the operator.

\subsection{Host-based anomaly detection}
Host-based anomaly detection systems are widely used in high-performance computing (HPC) clusters to localize faults.
They run on the host and continuously receive telemetry for analysis, without interacting with user applications.
Their main principle is to measure telemetry variability over time and detect outliers.
Prodigy~\cite{Prodigy10.1145/3581784.3607076} is an anomaly detection system that uses statistical features of a healthy telemetry dataset to train a variational auto-encoder (VAE).
If the VAE's reconstruction error for a sample exceeds a threshold, it is classified as an outlier.
Once detected, it uses an explanatory framework to report the attributes that triggered the alarm, to give engineers a starting point for debugging.
eACGM~\cite{xu2025eacgmnoninstrumentedperformancetracing} also uses telemetry data from the past hour to train a Gaussian mixture model (GMM).
Similarly, Greyhound~\cite{greyhound} uses Bayesian online change point detection on timeline data.
Unfortunately, this means a per-system healthy baseline is required for all those methods.
While this might be a reasonable assumption in HPC systems with homogeneous nodes, it does not hold in cloud environments because of the hardware and system diversity.
Furthermore, healthy baselines are costly to establish, difficult to maintain, and may not even exist.
VMs or containers can be launched on hosts with different hardware characteristics, making a uniform healthy baseline infeasible.
Moreover, performance variability is unavoidable and can be caused by operating system, software, library changes, or simply measurement error~\cite{222561}.
The unclear boundary between healthy and outlier makes the system prone to false positives~\cite{superbench24}.
As a result, systems such as Holmes~\cite{holmes} require retuning; Eroica~\cite{eroica} relies on intra-node discrepancies to detect anomalies.

\subsection{Case study}\label{sec:motivation-casestudy}

\begin{figure}[t]
    \centering
    \includegraphics[trim=70 165 80 160, clip, width=\columnwidth]{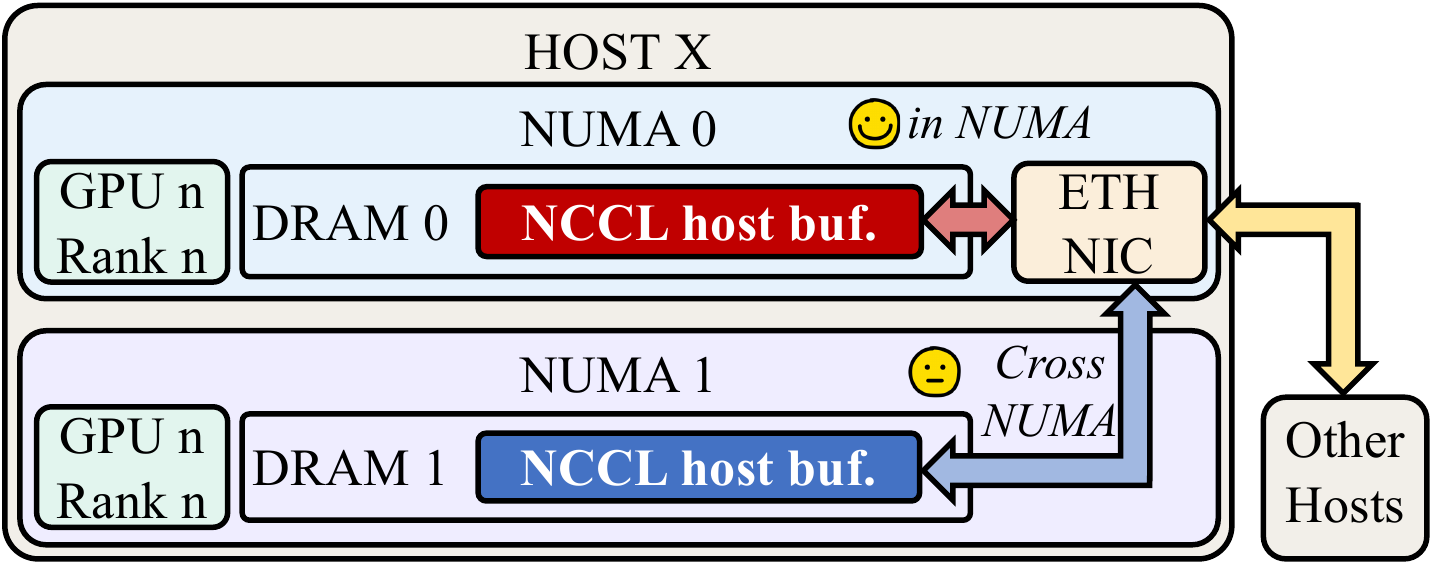}
    \vspace{-20pt}
    \caption{
    GPU--NIC NUMA placement (\autoref{sec:motivation-casestudy}).
    }
    \vspace{-20pt}
    \label{fig:numa_topology}
    \Description{Diagram of one host with two NUMA nodes. NUMA 0 keeps the NCCL host buffer beside the Ethernet NIC. NUMA 1 places that buffer on the other node, so traffic crosses NUMA before reaching the NIC and the other hosts.}
\end{figure}

We motivate our work using the example of a Non-Uniform Memory Access (NUMA) misplacement problem.
Despite being a simple misconfiguration issue, it is difficult to detect correctly with existing work.
This can happen on the cloud if the underlying hardware topology is hidden from the guest~\cite{10.1145/3725783.3764394}.
We perform supervised fine-tuning of a Llama-3.2-1B model~\cite{llama32modelcard} using the Alpaca dataset~\cite{alpaca} on a setup with two V100 GPUs (full configuration in \autoref{sec:evaluation}), deployed on VMs.
While cross-NUMA node communication is unavoidable because there is only one network interface card (NIC) and multiple GPUs, the critical communication path should be kept NUMA-local when possible.
For example, in \autoref{fig:numa_topology}, a communication process and buffer should be placed on NUMA-0.
A misplacement causes communication to traverse NUMA before reaching the NIC.
The inefficient configuration increases the average iteration time by 3.3\% and end-to-end runtime by 10.5\%.

\begin{figure}[t]
    \centering
    \includegraphics[trim=0 0 0 0, clip, width=0.9\linewidth]{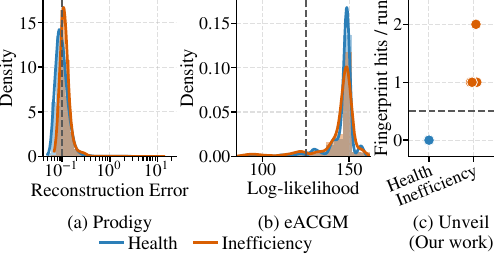}
    \vspace{-12pt}
    \caption{
    Comparison of Prodigy, eACGM, and \unveil (Our work) under GPU--NIC NUMA misalignment (\autoref{sec:motivation-casestudy}).
    }
    \vspace{-15pt}
    \label{fig:motivation-prodigy}
    \Description{Three panels for GPU--NIC NUMA misalignment. Prodigy reconstruction-error distributions for healthy and inefficient runs overlap, and many healthy samples sit above the alert threshold. eACGM log-likelihood distributions for the two runs are nearly the same. Unveil records fingerprint hits on inefficient runs and none on the healthy run.}
\end{figure}

We look at how Prodigy~\cite{Prodigy10.1145/3581784.3607076} and eACGM~\cite{xu2025eacgmnoninstrumentedperformancetracing}, which are the closest to \unveil, handle this issue.
Prodigy requires a healthy baseline, so we train its model with one healthy run, while eACGM uses unsupervised learning.
However, most application-related metrics are not available to eACGM due to being outside the VM.
After training, we restart the application and monitor it using eACGM and Prodigy.
We perform uninterfered runs (which is equivalent to the healthy run) and fault-injected runs, repeating each experiment five times.
Prodigy uses reconstruction error from its VAE to determine outliers.
We plot its frequency distribution in \autoref{fig:motivation-prodigy}(a).
An alert is generated when the error is above a 
threshold, denoted by the gray dotted line.
The distribution for inefficient runs is slightly shifted to the right, meaning it generated many alerts correctly.
However, more than 50\% of the error values from healthy runs are above the threshold, indicating a significant number of false positives.
Next, we look at eACGM, which uses a log-likelihood to generate alerts.
We plot the distribution between the two sets of runs in \autoref{fig:motivation-prodigy}(b).
Despite a clear difference in application performance, the log-likelihood distributions are indistinguishable.
This means eACGM detection also fails to distinguish the two scenarios.
Finally, we introduce our work \unveil in \autoref{fig:motivation-prodigy}(c), which not only successfully identified the uninterfered and fault-injected runs, but also accurately identified this as a NUMA misconfiguration issue.
Unlike systems such as Prodigy and eACGM, \unveil does not require a system-specific, dedicated healthy baseline run or full view of application metrics.
We explain this in detail in \autoref{sec:knowledge-base}.

\section{\unveil design}\label{sec:problemspace}

Our insight is that existing ML workload monitoring systems require application knowledge (e.g., iteration boundary information) or a healthy baseline.
After detecting an anomaly, they typically provide little guidance beyond isolating the affected node.
We design \unveil, an anomaly detection, diagnosis, and remediation system for ML in a cloud environment, with the following objectives.

\subsection{Objectives}

\noindent \textbf{Instrumentation-free and workload-agnostic.}
Existing work~\cite{Prodigy10.1145/3581784.3607076,xu2025eacgmnoninstrumentedperformancetracing,greyhound,holmes,xiao2026pulse,eroica,aegis,deng2025mycroft,superbench24}, even those that focus on operator-level deployment, require both the user and operator to work together.
We believe this is not suitable for a production cloud environment because it is unscalable and unreasonable. After all, only the cloud operators can apply system-level remediation.
\unveil should operate without any application knowledge.
For example, it must not be affected by unpredictable start/stop, while supporting diverse and unpredictable deployments, including training and inference, and across applications, such as computer vision (CV) and natural language processing (NLP).
Furthermore, \unveil supports local and distributed workloads, providing intra- and inter-node analysis.

\noindent \textbf{High portability.}
Given the complexity of the cloud technology stack, \unveil must be highly portable, without being coupled to any particular hardware, hypervisor, profiling tools, ML frameworks, job scheduling system, or being dependent on a specific kernel or OS.
To support wide systems with different hardware within a cloud, it must not require a separate baseline run for each setup.
Both \unveil's telemetry collection and detection pipeline should be easily expanded to handle new metrics, allowing it to adapt to new ML workloads, systems, and accelerators.

\noindent \textbf{Accurate and efficient detection.}
Storage and compute overhead challenge traditional application profiling. 
\unveil should achieve accurate detection with minimal metrics, ensuring low collection and analysis overhead for continuous deployment in production.

\noindent \textbf{Actionable diagnosis.}
Existing works identify only a faulty node, or pinpoint whether a problem is compute- or network-oriented.
Diagnosis still requires an offline benchmark or manual analysis.
Given the increasing scale of ML systems, it is infeasible to always rely on manual intervention.
Although automatic remediation is possible by evicting the faulty node through checkpoint-restart~\cite{greyhound}, it takes considerable time.
Therefore, it is important to identify possible automatic corrective actions, such as configuration changes.

\subsection{Design principles}\label{sec:framework}
We achieve the objectives with the following principles:

\noindent\textbf{Host-only operation.}
\unveil is deployed on the host and uses only host-visible telemetry.
We believe this is a suitable layer for monitoring modern cloud ML deployments, as it does not require instrumenting guest-side applications or libraries.
Application performance and inefficiency are also reflected in hardware signals.
As such, the same analysis pipeline works across different ML workloads, frameworks, schedulers, and hardware platforms.

\noindent \textbf{Within-run anomaly detection.}
ML workloads exhibit recurring execution phases during both training and inference~\cite{rhyner2024pimoptdemystifyingdistributedoptimization}.
For example, in recommendation model training, memory utilization steadily increases and then drops sharply as the workload repeatedly allocates and rebuilds large embedding tables.
This would generate false positives with change point detection algorithms.
\unveil uses unsupervised learning to exploit this regularity through within-run comparison and identifies normal behavior and anomalies.

\noindent\textbf{Accurate diagnosis and remediation.}
Accurate root cause diagnosis is difficult because no single metric can uniquely identify an inefficiency~\cite{superbench24}.
Systems such as eACGM train a probability model for \emph{each} metric (e.g., CUDA latency, Python latency), and directly attribute inefficiency to that category.
Prodigy operates as a black box and uses the CoMET~\cite{ates2021counterfactual} explanation framework to output the metrics that triggered an alert.
We believe performing anomaly detection and root cause attribution in a single step is unsuitable for diagnosis.
For example, an elevated TCP retransmission alone may indicate interrupt misplacement, congestion, or no problem at all.
When combined with CPU stalls and memory writeback pressure, it indicates network congestion.
However, when combined with a different communication anomaly, it can point to a completely different cause.
As such, \unveil deliberately performs detection and attribution in two distinct steps.
Information from detection is used to derive a \emph{signature}.
Signatures are qualitative and combine representative surrounding system behavior when an inefficiency exists, forming a portable negative baseline.
A database of known inefficiencies is used to identify and execute predefined remediation actions defined by engineers.

\noindent\textbf{Continuous deployment.}
Diagnosis must run continuously in production.
Framework profilers provide detailed execution traces, but they incur substantial runtime and storage overhead.
\unveil instead relies on a compact set of host-visible metrics.
This keeps telemetry collection and analysis lightweight. We quantify the resulting overhead in \autoref{sec:overhead}.

\noindent\textbf{Extensibility.}
Modern ML infrastructure is increasingly heterogeneous and evolves rapidly.
New accelerators, network devices, and system software continually introduce new telemetry interfaces.
Evolving ML workloads may also require new telemetry metrics.
\unveil therefore adopts a modular telemetry pipeline.
New probes, hardware platforms, and telemetry metrics can be incorporated independently with minimal implementation effort.

\subsection{Unsupervised anomaly detection}\label{sec:Methodology}

To enable host-only operation and accurate within-run anomaly detection, we adopt traditional anomaly detection methods widely used in AIOps~\cite{bodik2010fingerprinting,zhang2019crossdataset}.

\subsubsection{Metric Space Construction}\label{sec:metricspace}
Host-level monitoring exposes hundreds of telemetry signals (e.g., CPU core utilization, different types of cache misses, and network interface throughput), many of which are redundant.
For example, instructions, cycles, and IPC are strongly correlated.
Retaining all of them increases the sampling overhead, storage, and model dimensionality, which can lead to overfitting.
While systems such as Eroica~\cite{eroica} reduce storage cost by summarizing collected data, they do not reduce sampling overhead.
We instead identify redundant telemetry through correlation analysis.
\unveil defines an \emph{instance} as a monitored component, such as a CPU core or network interface.
A \emph{metric class} groups a measurement type shared across \emph{instances}.
Measuring a \emph{metric class} on an \emph{instance} produces a \emph{metric}.
For example, CPU utilization on core 0 is a \emph{metric} in the CPU core utilization \emph{metric class}.
To ensure good representation, we collect over 700 host-level \emph{metric}s from 30 representative ML workloads spanning NLP, large language model (LLM), and CV applications on both GPU and CPU-only clusters using standard tools (e.g., \texttt{perf}, \texttt{/proc}, and \texttt{nvidia-smi}).
Because execution time varies, we first align the time series using dynamic time warping (DTW) for comparison, then apply clustering using Pearson correlation.
Highly correlated metrics will be clustered together.
We rank the metrics within each cluster by prioritizing utilization-related ones, select the top metric, and discard the rest.
This is because they are easily interpretable.
The process reduces the number of metrics by 40\% (from approximately 700 to 400 per host), and lowers telemetry volume from approximately 23--37~KB/s to 14--22~KB/s while preserving coverage of major subsystems (full details in the supplementary material~\cite{chen2026supplementary}).

\subsubsection{Sliding Window Feature Extraction}\label{sec:tsfresh}
\unveil samples every 100ms to create a \emph{metric} vector (e.g., $<CPU0_{\%}, CPU1_{\%}, CPU_{LLC miss}, \allowbreak Throughput_{eth0}, ...>$).
Raw monitoring streams are noisy and autocorrelated, so individual samples provide limited information.
To capture temporal characteristics while reducing sensitivity to short-lived spikes, \unveil segments each metric's time series into overlapping 3~s windows with a 1~s stride.
We chose those parameters to preserve short-lived communication such as the sub-second all-reduce operations, and events such as transient GPU power transitions during kernel launches, while balancing overhead.
This window size is workload-agnostic because it is set based on the timescale of hardware-level events, rather than tuned to specific frameworks, workloads, or iteration lengths.
We show in \autoref{sec:evaluation} how it generalizes to other workloads without workload-specific recalibration.
At the end of each window, \unveil summarizes the samples by computing statistical and temporal \emph{feature}s, such as mean, variance, and autocorrelation for each \emph{metric}, which forms each \emph{metric}'s feature vector.
Since window processing may take longer than the stride (i.e., 1~s), \unveil avoids backlogs by processing windows in parallel with multiple workers.

\subsubsection{Online detection}
\label{sec:anomaly}
Production diagnosis systems commonly rely on manually tuned thresholds that must be reconfigured as workloads and deployment conditions change~\cite{onesizefitsnone24,Prodigy10.1145/3581784.3607076}.
These thresholds are difficult to generalize across infrastructure.
A conservative threshold reduces false positives, but severely delays detection.
For example, Mycroft reports a throughput anomaly only after bandwidth drops by 50\%~\cite{deng2025mycroft}.
At the same time, no single detector performs consistently across all workloads and anomaly types~\cite{superbench24,Prodigy10.1145/3581784.3607076}.
To overcome this, \unveil uses an ensemble of diverse models~\cite{xia2025nodesentry}.

\noindent \textbf{Z-score} measures the number of standard deviations a data point is away from a dataset's mean.
We compute the standard deviation of each feature relative to all previous windows, and flag a window that is above the $99^\text{th}$ percentile of the mean absolute Z-score~\cite{gitlab-prometheus-anomaly}.

\noindent \textbf{Mahalanobis distance} measures the distance between a data point and the center of a distribution~\cite{lakhina-sigcomm04,xu-sosp09}.
It is effective for recognizing joint patterns because it accounts for correlations and scale differences across metrics.
We project a feature vector onto a principal component that retains 95\% of the variance, then use it to compute the window's distance to the centroid of all previous windows.
We flag the distances above the $99^\text{th}$ percentile as anomalous.

\noindent \textbf{Isolation Forest} is a widely deployed anomaly detection method in production~\cite{aws-genai,aws-cloudwatch-rcf}.
It scales well, handles skewed and multi-modal feature distributions, while remaining robust to transient noise~\cite{aws-cloudwatch-rcf,he-iotbds2025,eyeri2023-anomaly}.
We train the model with a 1\% contamination rate, meaning we expect approximately 1\% of the data to be anomalous.

We choose these models to capture intra- and inter-metric behavior, as well as distribution- and non-distribution-based detection.
Z-score compares a data point against its own historical distribution only, whereas Mahalanobis distance takes into account other metrics' distributions jointly.
Isolation Forest distinguishes outliers by exploiting the fact that anomalous samples are easier to isolate than normal samples.
After computing the features of each metric at the end of a window, those feature vectors are concatenated and fed into the three models.

At the start of monitoring, the models initialize their references with the first five windows.
Thereafter, they independently evaluate each window against all the previous windows and run respective detections; the ensemble flags it as anomalous if at least two models agree.
This reduces model-specific false positives while covering diverse anomalies.
Any misconfiguration from the start becomes detectable when the workload begins to exercise the affected resource. An exception is if it was incorporated into the initial reference and remains completely stable throughout monitoring -- in which case the issue may remain undetected.

\subsection{Classifying inefficiencies}\label{sec:bottleneck}

Once anomalous windows are identified, \unveil moves on to identify the exact inefficiency and apply remediation.
Existing work such as SuperBench~\cite{superbench24} discussed the challenges of diagnosing subtle fail-slows that span multiple subsystems.
Identification requires running different combinations of microbenchmarks.
We follow their premise and posit that inefficiencies could be inferred using surrounding system behavior.

\begin{algorithm}[t]
\caption{Fingerprint Generation Using Anomalous Metrics}
\label{alg:fingerprint_generation}
\begin{algorithmic}[1]

\REQUIRE Sorted metric feature vector $V$
\ENSURE Set $S$ of metric classes with a list of feature names

\STATE $S \gets \emptyset$

\FOR{each metric feature $m \in V$}
    \STATE $c \gets \mathrm{MetricClassLabel}(m)$ \COMMENT{e.g., $CPU\%_{core0} \rightarrow CPU\%$}
    \STATE $f \gets \mathrm{FeatureName}(m)$ \COMMENT{e.g., mean, variance, slope}

    \IF{$c \in S$}
        \IF{$f \notin S[c]$}
            \STATE $S[c] \gets S[c] \cup f$ \COMMENT{Append the feature to the entry}
        \ENDIF
    \ELSE
        \STATE $S[c] \gets \{f\}$ \COMMENT{Add the new metric class to the set}
    \ENDIF

    \IF{$|S| = 3$}
        \STATE \textbf{break}
    \ENDIF

\ENDFOR

\STATE \textbf{return} $S$
\STATE \COMMENT{E.g., $S=\{CPU\%:\{\text{mean},\text{var}\}, \text{throughput}_{\text{eth0}}:\{\text{slope}\}, ...\}$}
\end{algorithmic}
\end{algorithm}

\subsubsection{Fingerprinting}\label{sec:fingerprint-generation}

We develop a \emph{fingerprint} extraction algorithm to represent inefficiencies.
We first use the detection pipeline in \autoref{sec:anomaly} to identify anomalous windows.
Each anomalous window has a vector containing all the feature values for all metrics.
After that, we sort the vector elements by their z-scores.
This is because z-score is straightforward and rankable.
We maintain an attribute mapping, so we can reverse-lookup a value's label later.

We describe fingerprint generation in Algorithm~\autoref {alg:fingerprint_generation}.
Each set entry groups a \emph{metric class} with one or more \emph{feature} names.
Initially, a \emph{metric class} (MetricClassLabel), together with a \emph{feature} name (e.g., mean), will be added to a set.
We move to the next element in the vector.
If this element has the same \emph{metric class}, for example, CPU utilization on another core, we perform an update of its \emph{feature} name.
If this is the same kind of \emph{feature} (e.g., also a mean), we move on to the next element.
Otherwise (e.g., a slope), we append it to the CPU utilization entry in the set.
If the next element's \emph{metric class} is not in the set (e.g., network throughput), we add it to the set.
This process repeats until we have three entries in the set.
One example result is shown in L17 of Algorithm~\autoref{alg:fingerprint_generation}.

Our greedy selection algorithm is motivated by three principles.
Firstly, to capture a wide variety of \emph{metric classes}, not only those representing different \emph{instances}.
For example, we are concerned with a CPU core behaving anomalously, rather than which specific core it is.
Secondly, to capture important intra-class variations, e.g., if another CPU core behaved anomalously in a different way (i.e., through a different statistical feature), it should also be taken into account, without sacrificing class coverage.
Thirdly, we want a portable, qualitative fingerprint to avoid any fixed quantitative baseline.
Thus, we drop the numerical features and retain the labels.

\subsubsection{Inefficiency signatures}\label{sec:fingerprint-matching}

\begin{figure}[t]
    \centering
    \includegraphics[trim=9 16 0 3, clip, width=\columnwidth]{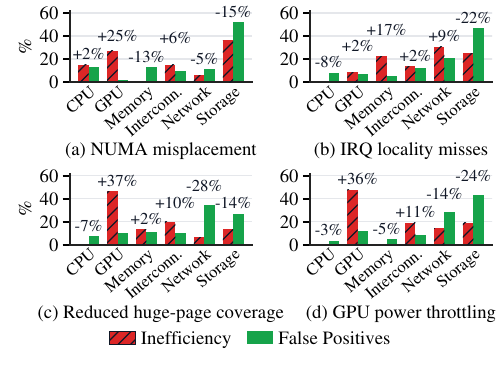}
    \vspace{-20pt}
    \caption{Distribution of fingerprints at the subsystem-level for four silent inefficiency types (panels (a)--(d), red). Red bars denote fingerprints observed under the injected inefficiencies, while green bars denote fingerprints from anomalous windows detected during healthy executions ("False Positives") (\autoref{sec:fingerprint-matching}).}
    \vspace{-20pt}
    \label{fig:fingerprint-shift}
    \Description{Four bar charts of fingerprint share across CPU, GPU, memory, interconnect, network, and storage. Panels are NUMA misplacement, IRQ locality misses, reduced huge-page coverage, and GPU power throttling. Red bars are injected inefficiencies and green bars are false positives from healthy runs. The red bars concentrate on different subsystems from the green bars.}
\end{figure}

We generate fingerprints for the anomalous windows from the runs with and without injected inefficiencies.
Despite the robustness of our detection, false positives are inevitable due to system noise.
What makes them different from genuine inefficiencies is the systematic behavior.
As such, fingerprints from true positives should differ distinctly and consistently from false positives.
We validate this assumption using the following frequency analysis.
For each fingerprint of a given inefficiency, we map its three \emph{metric classes} to the following subsystems: CPU, GPU, memory, network, interconnect, and storage, and normalize their frequencies as percentages.
After that, we arrive at a ratio of where inefficiencies arise between the six subsystems.
Using the healthy runs as a baseline, we show how the inefficiencies differ in \autoref{fig:fingerprint-shift}.
For example, \autoref{fig:fingerprint-shift}(a) shows a communication process that is placed on an incorrect NUMA node.
The frequency distributions differ significantly, where GPU-related anomalies are overrepresented, whereas storage-related ones are underrepresented.
A naive diagnosis would classify this inefficiency as having GPU problems.
It was not true.
However, by combining three \emph{metric classes} and associated features in the fingerprint, we can narrow this down to a NUMA misplacement.
This motivates us to build a knowledge base of known inefficiencies to record those fingerprints.

\subsection{Runtime remediation}\label{sec:auto-remediation}
We build a portable knowledge base with known inefficiencies and recipes to address them, if possible without interrupting running applications.

\subsubsection{Knowledge base}\label{sec:knowledge-base}
As shown in \autoref{fig:fingerprint-shift}, different inefficiencies produce different fingerprint distributions.
While Aegis~\cite{aegis} and SuperBench~\cite{superbench24} also use benchmarking to localize faults, our approach attempts to \emph{run-once-work-everywhere}.
We first perform an end-to-end \emph{baseline run} with microbenchmarks, without injecting any inefficiencies, on a cluster with V100 GPUs (detailed configuration in \autoref{sec:evaluation}).
After that, for each inefficiency, we perform a \emph{faulty run} with controlled fault injection.
Finally, we compare the signatures of detected windows from the \emph{faulty run} against the \emph{baseline run}, and discard those that appear in both as false positives.
Those fingerprints qualitatively describe systematic behavior while an inefficiency is present, making them portable across systems.
Notably, the \emph{baseline run} is only for knowledge base development, and not for runtime detection and matching.

\paragraph{NUMA misplacement}\label{sec:numa-microbench}
We run the NCCL send/recv microbenchmark with 256\,MB messages in a case where the communication process has been placed on a NUMA node away from the NIC.
This results in excessive cross-NUMA data movement, resulting in significantly lower throughput (approximately 40\% with TCP and 5\% with InfiniBand RDMA).
Although NUMA placement is a well-known issue, identification is difficult because the operator only observes reduced communication performance rather than placement errors themselves.
Furthermore, cloud virtualization hides underlying hardware topology, leading to the application's autoconfiguration failure~\cite{10.1145/3725783.3764394}.

Two fingerprints recur across inefficient runs.
The first combines abnormal periodicity in GPU memory clocks, change in GPU PCIe link generation, and trend in CPU frequency.
The second combines abnormal stationarity in storage I/O latency, stationarity in storage write latency, and variance in GPU power.
Interestingly, while the root cause of NUMA misplacement is on the CPU and memory side, the fingerprints appear across GPU and storage subsystems.
This suggests that the local placement error spreads pressure through the communication path and causes a cross-layer anomaly.
This demonstrates the importance of cross-subsystem attribution.
We add both fingerprints to the knowledge base.

The misplacement occurs on the host, which can be remediated by moving the process using \texttt{taskset(1)} and migrating its memory pages using \texttt{migratepages(8)}.
If Direct Memory Access (DMA) memory pinning prevents page migration, the application needs to be interrupted.
This can be coordinated using a guest notification mechanism to restart the application with the NCCL proxy thread rebound to the NIC-local NUMA node, either via \texttt{numactl} or by supplying NCCL with an accurate topology file~\cite{aegis}.

\paragraph{IRQ locality misses}\label{sec:irq-microbench}
NIC interrupts are handled by interrupt request (IRQ) threads on the CPU. We again run the NCCL send/recv microbenchmark with 256 MB messages in a case where the IRQ thread is placed on a NUMA node away from the NIC. If an application process that uses that data, such as a TCP client, is not placed near the IRQ thread, it triggers cross-NUMA communication and reduces throughput by 32\%.

Three fingerprints recur.
The first combines abnormal stationarity in GPU PCIe link generation, stationarity in memory writeback, and trend in storage I/O latency.
The second combines abnormal stationarity in TCP fast retransmissions, storage I/O latency, and TCP retransmissions.
The third is characterized by a combination of abnormal variability in swap cache, abnormal variability in TCP SYN retransmissions, and low GPU power.
We add all three fingerprints to the knowledge base.
These fingerprints differ from those produced by NUMA misplacement.
In this case, TCP-related signals show clear abnormal behavior directly.
This is consistent with the fact that IRQ thread placement affects the network data path more directly.
We detect this case within an average of 53.6s, and it can be corrected without interrupting running applications, by re-pinning the affected IRQ handlers to the NUMA node local to the NIC through NIC-vendor-provided binding scripts.

\paragraph{Reduced huge-page coverage}\label{sec:memoryfrag}
In long-running applications, memory fragmentation gradually reduces the availability of contiguous 2~MB pages, lowering the success rate of transparent huge-page (THP). This forces fallback to 4~KB pages, increasing translation lookaside buffer (TLB) pressure and address translation overhead. We develop and run a recommendation-style inference microbenchmark with 26 embedding tables totaling 133GB. Each query performs batched sparse CPU embedding lookups across all tables (10M$\times$128 each, DLRM-style), mirroring production recommendation serving. We emulate this effect by setting the Linux THP mode to \texttt{always}, which never forces allocations to use small pages, and \texttt{never}, which only uses small pages. This causes throughput to drop from 184K to 112K queries/s ($1.64\times$ degradation), and a $3.4\times$ increase in TLB-load misses.

Three fingerprints recur.
The first combines abnormal activity in GPU memory clocks, variation in GPU power, and stationarity in PCIe link generation.
The second combines a high skewness in network receive-drop measurements, with abnormal variation in storage writes and minimum in memory writeback.
The third combines abnormal periodicity in unevictable memory, abnormal periodicity in GPU power, and trend in storage I/O time.
We add all three fingerprints to the knowledge base.
\unveil can detect this in an average of 164.0s. 
Memory fragmentation is host-side and can be remediated by defragmenting memory using Linux's \texttt{khugepaged} compaction.
Running \texttt{khugepaged} does not interrupt the application, but a workload restart can improve defragmentation.

\begin{figure}[t]
    \centering
    \includegraphics[width=\columnwidth]{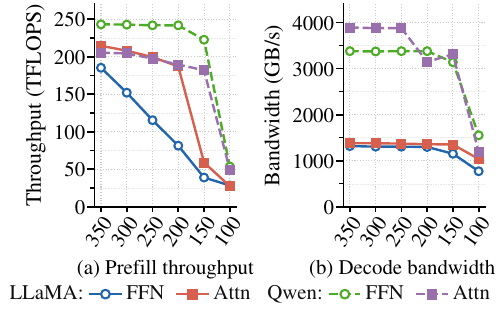}
    \vspace{-25pt}
    \caption{Impact of GPU power limits on prefill throughput and decode bandwidth (\autoref{sec:gpu-power-microbench}).}
    \label{fig:gpu-power-prefill}
    \Description{Line charts of LLaMA and Qwen feed-forward and attention layers as the GPU power limit falls from 350 W to 100 W. Prefill throughput drops earlier and more steeply than decode bandwidth. LLaMA feed-forward throughput declines almost linearly, while the other series stay flat until a sharp drop at the lowest power limits.}
    \vspace{-20pt}
\end{figure}

\paragraph{GPU power throttling}\label{sec:gpu-power-microbench}
Reduced GPU power or streaming multiprocessor (SM) frequency does not always affect application performance. To characterize when performance is affected, we introduce gradual GPU power limits for LLM workloads in \autoref{fig:gpu-power-prefill}. Analysis shows that sensitivity to power reduction varies across applications. For example, the throughput of LLaMA-FFN prefill degrades linearly, whereas other workloads show a sharp drop only under substantial power reduction. The effect is stronger on prefill than on decode (\autoref{fig:gpu-power-prefill}(b)). This is expected, since prefill is compute-intensive while decode is more memory-intensive.
\unveil detected the anomalous windows in \autoref{fig:gpu-power-prefill} with an average delay of 34.8s. 

Two fingerprints recur.
The first combines abnormal temporal dynamics in storage write latency, activity in GPU memory clocks, and periodicity in network traffic. 
The second combines abnormal variation in PCIe link generation, burstiness in GPU power, and trend in GPU memory clocks.  
We add both fingerprints to the knowledge base.
Using these fingerprints, \unveil can identify anomalous windows where performance degrades due to power fluctuations. 
This can be remediated by redistributing per-GPU power limits to reduce local heat density, or by adjusting the cooling system configuration through the system's software management interface.

\subsubsection{Signature matching and remediation}\label{sec:automatic-patching}
Our fingerprints consist of text labels of three \emph{metric classes}, each containing a set of \emph{features}.
\unveil compares the anomalous window's fingerprint with the ones in the knowledge base.
We first compare each \emph{metric class} that appears in both the query and the candidate fingerprint.
If they share at least one \emph{feature}, the \emph{metric class} is matched.
A candidate fingerprint is matched if at least two metrics are matched.
Our analysis (\autoref{fig:fingerprint-shift}) shows that one of the metrics in a fingerprint tends to dominate, followed by two less dominant ones that reflect surrounding system behavior.
Therefore, a two-of-three rule takes into account system variability, while preserving cross-subsystem behavior.
We note that this is not a confidence score.
If there are multiple matches, \unveil selects the one with the highest similarity and executes its associated remediation scripts.
In case of a tie, all corresponding scripts are executed.
If an inefficiency requires manual intervention, operators can instead use this script to gather extra data or perform steps that aid manual remediation later.

\section{Framework and implementation}\label{sec:implementation}

\begin{figure}[t]
    \centering
    \includegraphics[trim=160 195 340 175, clip, width=\columnwidth]{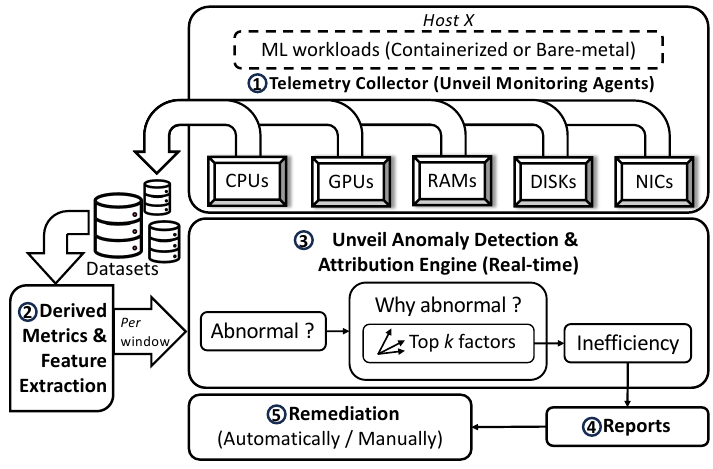}
    \vspace{-24pt}
    \caption{Hardware-centric observability and remediation workflow in \unveil (\autoref{sec:implementation}).}
    \vspace{-15pt}
    \label{fig:framework}
    \Description{Flowchart of Unveil on one host. Monitoring agents collect telemetry from CPUs, GPUs, memory, disks, and NICs while containerized or bare-metal workloads run. Each window is turned into features, tested for an anomaly, and attributed to the top contributing factors. The result is reported and can be remediated automatically or manually.}
\end{figure}

We implement \unveil as an immediately deployable real-time detection framework that runs as a lightweight background daemon.
The framework consists of a pipeline illustrated in \autoref{fig:framework}.

\paragraph{Telemetry register}
\unveil exposes an interface to define how tools are executed and how output is parsed.
When defining output parsing, the user can optionally generate new \emph{derived metrics}.
For example, the user can create an IPC metric using the instructions and CPU cycles reported by \texttt{perf}.
As such, new profiling tools can be easily registered, e.g., when new accelerators are deployed.

\paragraph{Metric analysis engine}
Once all the probes are registered, \unveil performs the metric selection analysis again in \autoref{sec:metricspace}.
However, \unveil already works out-of-the-box with common tools such as \texttt{perf}~\cite{linux-perf}, \texttt{procfs}~\cite{procfs}, and \texttt{nvidia-smi}~\cite{nvidia-smi-docs}.
Therefore, rerunning the analysis is not necessary when using \unveil in a standard deployment.

\paragraph{Knowledge base}
\unveil maintains a knowledge base as a simple key-value store, where fingerprints are used as keys and remediation scripts are stored as values.
We have created and preloaded the knowledge base with well-known inefficiencies
(\autoref{sec:knowledge-base}, full list in the supplementary material ~\cite{chen2026supplementary}) which can be expanded by cloud engineers.
They review unmatched fingerprints and classify them as false positives, new inefficiencies, or new signatures of known inefficiencies.
For new inefficiencies, they develop and validate remediation scripts before adding them to the knowledge base.

\paragraph{Telemetry collector and time series database}
The telemetry collector periodically (100ms by default) uses the probes in the telemetry register to sample metrics.
The probes sample in parallel; their outputs are parsed using predefined instructions, time-aligned, and stored in a time series database as one record.
The collector maintains the sliding windows (\autoref{sec:tsfresh}) and calculates statistical features once a window is full.
The feature vector of a window is sent for anomaly detection.
The collector can be turned on by the operator at any time they want to use \unveil.

\paragraph{Detection and remediation}
The three models evaluate each window's feature vector independently (\autoref{sec:anomaly}).
If at least two flag a window, \unveil calculates its fingerprint (Algorithm~\autoref{alg:fingerprint_generation}) and queries the knowledge base (\autoref{sec:knowledge-base}).
Once matched, \unveil automatically runs the associated remediation script (\autoref{sec:automatic-patching}).
If a window is detected but unmatched, we report it in a log entry for engineer review and proceed.

\paragraph{Reporting}
\unveil periodically generates intra- and inter-host reports, which summarize resolved or unresolved inefficiencies.
In inter-host reports, telemetry from many hosts is aggregated to locate deviations.
Using the report and the time series data, engineers can investigate false positives and incorporate new inefficiencies into the knowledge base.

\unveil is implemented in Bash and Python, with 8.7k LoC.
We use the \texttt{tsfresh} Python package to perform time series analysis and feature extraction.
\texttt{scikit-learn} and \texttt{numpy} are used to develop our metric space analysis engine and anomaly detection models.
Apache Parquet~\cite{vohra2016apache} is used for storing time series data.

\section{Evaluation}\label{sec:evaluation}
We evaluate \unveil on a single-node system and two GPU clusters to demonstrate its low overhead, detection performance, and effectiveness on realistic, unseen end-to-end workloads.
Finally, we position \unveil against two state-of-the-art operator-level monitoring systems, Prodigy~\cite{Prodigy10.1145/3581784.3607076} and eACGM~\cite{xu2025eacgmnoninstrumentedperformancetracing}. 

\subsection{Experimental Setup}\label{sec:setup}

\noindent\textbf{Hardware Platforms.}
We evaluate \unveil on three systems. 
\begin{itemize}[leftmargin=*]
    \item \textbf{V100} consists of two dual-V100 nodes (4 GPUs total) with $2\times$ Intel Xeon Platinum 8268 24-core CPUs (48 cores total) and 377~GB RAM, connected by HDR100 InfiniBand and 10\,GbE.
    \item \textbf{L40S} consists of four quad-L40S nodes (16 GPUs total) with $2\times$ Intel Xeon Gold 6548N 32-core CPUs (64 cores total) and 500~GB RAM, connected by NDR400 InfiniBand and 100\,GbE.
    \item \textbf{ADA} is a single-node, non-NUMA system on which we have full root privileges. It has $2\times$ NVIDIA RTX4000 Ada Generation GPUs (20~GB each), one AMD Ryzen Threadripper~PRO5955WX 16-core CPU, and 128~GB RAM, connected by 1\,GbE.
\end{itemize}
We do not include a production cloud environment in this evaluation to facilitate safe and controlled fault injection experiments, but the underlying principles remain valid.

\noindent\textbf{System environment.}
All experiments run Python~3.10, CUDA~12.6, and PyTorch~2.6 with NCCL configured for the underlying InfiniBand or TCP fabric.
We run ML applications using frameworks such as Megatron~\cite{megatron-lm}.
The applications are deployed through unprivileged Singularity~\cite{singularity} containers.

\noindent\textbf{Knowledge-base preparation.}
We curated the inefficiency signatures in \autoref{sec:knowledge-base}, manually developed and validated the respective remediation scripts before evaluation.

\noindent More details of the experimental setup are provided in the supplementary material~\cite{chen2026supplementary}.

\subsection{Cost and Efficiency}\label{sec:overhead}
We first evaluate how \unveil achieves its efficient detection objective against related work.
We run a full 3B-parameter GPT-style LLM fine-tuning workload on ADA with $\approx$52k training samples using mixed parallelism, which takes $\approx$9--12 minutes.
We use ADA for this experiment in order to deploy eACGM, which requires root privileges that we do not have on the clusters.
Since \unveil runs independently on every host, the single-node system analysis applies even when deployed at large scale.

\subsubsection{Monitoring overhead}\label{sec:monitoring-overhead}

\begin{figure}[t]
    \centering
    \includegraphics[trim=0 0 0 0, clip, width=\columnwidth]{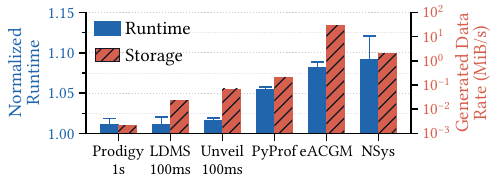}
    \vspace{-25pt}
    \caption{Runtime overhead (left axis) and telemetry data rate (right axis) across relevant tools (\autoref{sec:monitoring-overhead}).}
    \label{fig:overhead-comparison}
    \Description{Grouped bars for Prodigy at 1 second, LDMS at 100 ms, Unveil at 100 ms, PyProf, eACGM, and Nsight Systems. Blue bars are normalized runtime and stay near one for Prodigy, LDMS, and Unveil, while the profilers are higher. Hatched bars are the log-scale telemetry rate, lowest for Prodigy and Unveil and highest for eACGM.}
\end{figure}

We show the normalized results in \autoref{fig:overhead-comparison}, which compares the end-to-end time with and without monitoring using the blue bars and left-Y axis.
\unveil, which samples at the default 100ms rate, introduces significantly lower overhead ($1.01\times$) than profilers such as Nsight Systems ($1.09\times$) and PyProf ($1.05\times$).
This is because \unveil does not trace or interact with the applications.
Interestingly, eACGM ($1.08\times$) performs similarly to Nsight Systems.
Although eACGM monitors in the kernel, it still performs application-level tracing through eBPF entry points (e.g., \texttt{PyObject\_CallFunction}).
This results in a similar overhead to other application-level profilers, while requiring root privilege.
\unveil is only narrowly beaten by Prodigy ($1.01\times$), which also uses commonly available tools like \unveil to sample periodically.
This is expected, because Prodigy only samples once per second.
To perform a comparison, we use LDMS~\cite{osti_1106397}, the telemetry collector of Prodigy, and sample all supported metrics every 100~ms, resulting in comparable execution time to \unveil.

We now look at the storage overhead (red bars and right-Y axis in \autoref{fig:overhead-comparison}, in log scale).
Unsurprisingly, profilers generate the most data. PyProf, Nsight Systems, and eACGM generate up to 29.4~MiB/s, whereas \unveil generates 0.07~MiB/s.
This means that profilers can generate up to 2,482~GB of data per day, compared to 6~GB by \unveil, showing \unveil is suitable for continuous deployment.
Prodigy's default model only uses 156 metrics and generates 0.002~MiB/s, unlike \unveil, which samples over 400 metrics.
When sampling all supported metrics with LDMS every 100ms, it generates 0.02~MiB/s, which is comparable to \unveil.

\subsubsection{Monitoring cost across sampling frequencies}\label{sec:overhead-sample-frequency}
\begin{figure}[t]
    \centering
    \includegraphics[trim=0 10 0 0, clip, width=\columnwidth]{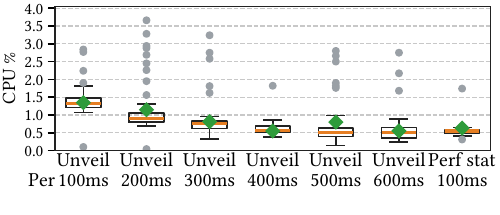}
    \vspace{-24pt}
    \caption{CPU overhead of \unveil at different sampling intervals, compared with \emph{perf stat} at 100~ms (\autoref{sec:overhead-sample-frequency}).}
    \label{fig:tool_overhead}
    \Description{Box plots of CPU utilization for Unveil at sampling intervals from 100 ms to 600 ms, next to perf stat at 100 ms. The median falls from about 1.3 percent at 100 ms toward about 0.5 percent at longer intervals, close to perf stat. A few samples exceed 2 percent.}
\end{figure}

We next measure the cost of \unveil's telemetry collection at different sampling frequencies. \unveil runs on an empty ADA system, and average CPU utilization is monitored. For comparison, \texttt{perf stat} is configured with a 100~ms collection interval.
\texttt{perf} is lightweight and only adds approximately 0.5\% average CPU usage.
At the default sampling rate (i.e., 100ms), \unveil adds approximately 1.3\%.
It is roughly equivalent to 2.6 \texttt{perf stat} processes.
If we increase the interval to 600ms, it is approximately the same as running \texttt{perf}.
This keeps telemetry collection overhead low.

\subsubsection{Detection cost}
After showing \unveil's monitoring adds little overhead to applications, we now quantify the cost of anomaly detection and analysis using 3~s windows, a 1~s stride, and three workers in parallel.
We break down the absolute cost of each step in the pipeline.
It takes approximately 2.26~s to process each sliding window, of which 1.46~s is spent calculating statistical features, and 0.667~s is spent on other operations.
Running the unsupervised detection models takes only up to 0.017s.

\subsection{Detection performance}
Next, we evaluate how \unveil achieves accurate detection objectives by reusing the same workload in \autoref{sec:overhead}. We use the L40S system because it was unseen by \unveil's metric analysis and knowledge base before (which was created on V100).

\subsubsection{False Positives}\label{sec:false-positives}

\begin{figure}[t]
    \centering
    \includegraphics[width=\columnwidth]{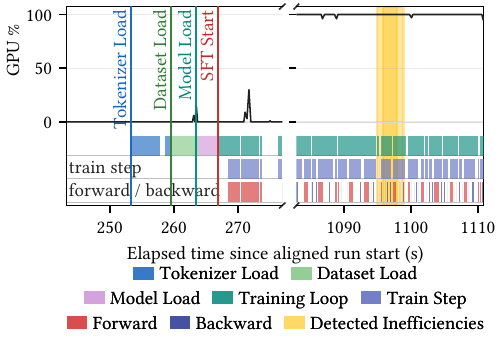}
    \vspace{-25pt}
    \caption{NVTX-annotated execution timeline. Detected inefficiencies (yellow) rarely coincide with lifecycle transition boundaries (\autoref{sec:false-positives}).}
    \label{fig:nvtx-alignment}
    \Description{Timeline of GPU utilization and NVTX phases, including tokenizer load, dataset load, model load, supervised fine-tuning start, training steps, and forward or backward. Yellow bands mark detected inefficiency windows. Those bands rarely line up with the vertical lines at phase changes.}
    \vspace{-15pt}
\end{figure}

Outlier detection that relies on individual metrics and trained baselines is prone to false positives~\cite{superbench24,deng2025mycroft}.
Unveil overcomes this by using an ensemble of unsupervised models (\autoref{sec:anomaly}) and a sliding window (\autoref{sec:tsfresh}).
A major reason for false positives is phase transitions within an application, such as between encoding and decoding, or data ingestion and compute.
This causes natural changes in hardware utilization and deviates from recent history.
We examine whether \unveil is robust to false positives by validating against NVIDIA Tools Extension (NVTX)~\cite{nvidiaNVTXNVIDIA} labels.
We rerun the fine-tuning workload and inject NVTX labels at change points to create an annotated timeline using Nsight Systems.
Then, we use \unveil to detect anomalous windows.
\autoref{fig:nvtx-alignment} shows a part of the workload; blocks denote workload phases, and lines denote change points; yellow windows mark detected windows.
After multiple runs, our analysis shows that only 1\% of detected windows overlap with annotated phase transitions, and none match a known inefficiency signature.
As a result, these overlapping anomalies do not trigger remediation. 
While the detectors do not explicitly recognize phase transitions, subsequent windows update the models' within-run reference, which gradually reflects the new phase.

\subsubsection{Ensemble detection and fingerprinting accuracy}
\label{sec:detector-selection}

\begin{table}[t]
\centering
\small
\setlength{\tabcolsep}{3pt}
\caption{Detection and fingerprint matching performance between different combinations. Z = Z-score, IF = Isolation Forest, and Maha = Mahalanobis distance (\autoref{sec:detector-selection}).}
\vspace{-10pt}
\label{tab:detector_perf}
\begin{tabular}{lcccc}
\toprule
\textbf{Configuration} &
\shortstack{\textbf{Signature}\\\textbf{Hit Rate}} &
\shortstack{\textbf{Signature}\\\textbf{Coverage}} &
\shortstack{\textbf{Detection--}\\\textbf{Signature}\\\textbf{Agreement}} &
\shortstack{\textbf{Healthy}\\\textbf{Contamination}} \\
\midrule
$Z \lor IF \lor Maha$ & 6.1\% & \textbf{81.3\%} & 11.2\% & 3.3\% \\
$Z \lor IF$           & 6.1\% & \textbf{81.3\%} & 11.2\% & 3.3\% \\
$Z \lor Maha$         & 6.2\% & \textbf{81.3\%} & 11.2\% & 3.2\% \\
$IF \lor Maha$        & 37.2\% & 51.2\% & 40.6\% & 1.2\% \\
$Z$                   & 6.2\% & \textbf{81.3\%} & 11.2\% & 3.2\% \\
$IF$                  & 41.6\% & 39.5\% & 37.8\% & 2.3\% \\
$Maha$                & 23.4\% & 15.6\% & 18.0\% & \textbf{0\%} \\
$Z \land IF$          & \textbf{46.4\%} & 39.5\% & 39.3\% & \textbf{0\%} \\
$Z \land Maha$        & 23.4\% & 15.6\% & 18.0\% & \textbf{0\%} \\
$IF \land Maha$       & 21.3\% & 3.9\% & 8.5\% & \textbf{0\%} \\
$Z \land IF \land Maha$ & 21.3\% & 3.9\% & 8.5\% & \textbf{0\%} \\
\textbf{2/3 vote} & \textbf{40.0\%} & \textbf{51.2\%} & \textbf{41.8\%} & \textbf{0\%} \\
\bottomrule
\end{tabular}
\end{table}

We now evaluate the performance of different detector combinations in \autoref{tab:detector_perf}.
We run the detectors on telemetry from both healthy and fault-injected inefficiency.
For evaluation, we generate and match fingerprints for every window, even when not flagged as anomalous.

On injected runs, let $N_{\mathrm{hit}}$ denote the number of detected windows matching the expected signature.
\emph{Signature hit rate} is $N_{\mathrm{hit}}$ divided by all detected windows, while \emph{signature coverage} is $N_{\mathrm{hit}}$ divided by all windows matching the expected signature.
\emph{Detection--signature agreement} is their macro-averaged harmonic mean.
On healthy runs, contamination is the number of detected windows matching any known signature divided by all detected windows.
An injected inefficiency may not affect all windows in a run because slowdowns depend on resource use and runtime variability.
These measures therefore evaluate detection and fingerprint matching, not diagnosis accuracy against independent window-level ground truth.
Our configuration with a two-thirds majority vote achieves the highest fingerprint \emph{Detection--signature agreement} (41.8\%).
Our results show that \unveil successfully covers all the fault types in the knowledge base.

\begin{figure}[t]
    \vspace{-13pt}
    \centering
    \includegraphics[width=\columnwidth]{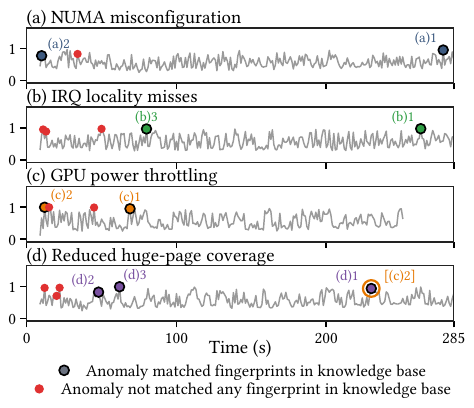}
    \vspace{-27pt}
    \caption{Detected anomalies (red markers) and knowledge-base fingerprint matches (black-ringed markers) over time (\autoref{sec:detector-selection}).}
    \vspace{-20pt}
    \label{fig:casestudy-finterprint}
    \Description{Four anomaly-score traces over 285 seconds for NUMA misconfiguration, IRQ locality misses, GPU power throttling, and reduced huge-page coverage. Colored circles are windows matched to a knowledge-base fingerprint. Red circles are detected but unmatched. One huge-page window is also marked as a GPU-power fingerprint.}
\end{figure}

Next, we run the same microbenchmarks used to generate the knowledge base in \autoref{sec:numa-microbench} on L40S.
As this is a new environment, it evaluates the portability of the knowledge base (generated on V100) and anomaly detection without a new healthy baseline run.
\autoref{fig:casestudy-finterprint}(a)--(d) plot the results, where red circles denote anomalous windows that are not matched, and colored circles denote matching in the knowledge base.
The Y-axis shows the anomaly score, which is a normalized average of the values reported by the three detection models.
\unveil successfully identified the anomalous windows in all four cases, and matched some of them to the correct inefficiency in the knowledge base.
Unmatched windows do not trigger remediation and are logged for analysis.
We have only detected one false identification, where a window in the reduced large-page coverage benchmark matches the fingerprint of GPU power throttling.
This reflects the need to expand and continuously evolve the knowledge base to fill the fingerprint space.

\subsubsection{Comparison with Existing Approaches}

We illustrated in \autoref{sec:motivation-casestudy} and \autoref{fig:motivation-prodigy} how Prodigy and eACGM fail to distinguish inefficient environments from healthy ones.
We further examine their performance using the microbenchmark and controlled degradation experiments in \autoref{sec:detector-selection}.
On top of the NUMA misplacement case, Prodigy fails to detect reduced GPU headroom cases, even when trained on healthy baseline executions.
eACGM detects these inefficiencies only when deployed inside the guest, where CUDA, PyTorch, and NCCL runtime events are available.
When deployed from the host, this runtime context is unavailable.
As a result, eACGM fails to detect all four silent inefficiencies.
Unlike eACGM, \unveil relies only on host-visible telemetry.

\subsection{End-to-end case studies}
\label{sec:real-world}

Finally, we apply \unveil to two applications on the L40S system to demonstrate its portability and actionable diagnosis. After remediation, we rerun the application to verify that the corresponding matched fingerprints do not recur over a complete execution.
These reruns are for the evaluation, not part of the remediation process.

\subsubsection{GPU power throttling}
We examine a case of GPU power throttling in an LLaMA-3.1-70B-Instruct~\cite{llama31modelcard} (LLaMA) serving workload using tensor parallelism, with requests served sequentially.
The cluster suffers from severe GPU power throttling (to 100~W).
Once throttling occurs, throughput decreases by 7\% to 31.4~token/s, and p95 time-to-first-token (TTFT) increases to 857~ms (2.3$\times$ increase).

\unveil detects the anomaly within 33.6 seconds.
Specifically, the matched signatures jointly capture statistical characteristics of GPU power, including variance, autocorrelation, and ADF, together with storage write activity, including the variance, absolute energy, ADF, and temporal trends of write counts, written sectors, and I/O service time.
The co-occurrence of these characteristics enables \unveil to robustly identify the GPU throttling fingerprint under production LLM serving workloads and correctly attribute the detected anomalies to reduced GPU compute capability.
\unveil matched its signature and automatically executed the relevant remediation script.
The signature disappeared after GPU power correction.

\subsubsection{Memory fragmentation}
We examine LLM serving with Qwen3-235B-A22B~\cite{qwen3technicalreport} (Qwen), whose continuous-batching workload overlaps prefill and decode execution.
When huge-page coverage is reduced, p95 TTFT increases by $1.7\times$, while throughput and time per output token (TPOT) remain unchanged.
This is consistent with increased TLB pressure disproportionately affecting prefills.

\unveil detects the degradation approximately 31.0~s after start.
It detected several anomalous windows, some of which are matched with the huge-page coverage inefficiency in the knowledge base.
The matched fingerprints jointly capture statistical characteristics of memory activity, storage I/O, and network behavior, including the variance, absolute energy, quantiles, and temporal trends of writeback traffic, storage write activity, and I/O service time.
The matched fingerprints disappeared after memory defragmentation using the associated remediation script.

\subsubsection{When Inefficiencies Do Not Manifest}
We further examine another two inefficiencies included in the knowledge base: NUMA misconfiguration and IRQ locality misses.
Under the evaluated LLaMA and Qwen serving deployments, neither inefficiency produces measurable degradation in TTFT, TPOT, or throughput.
Correspondingly, the observed fingerprints from detected anomalous windows do not match the NUMA or IRQ signatures.

This result provides two insights.
First, production LLM serving frameworks may absorb or hide low-level inefficiencies through their execution pipeline, scheduling strategy, or communication pattern, preventing these inefficiencies from propagating to end-to-end performance.
Second, because these inefficiencies do not manifest as performance anomalies, their signatures are not activated during attribution.
This is desirable because it indicates that \unveil attributes only inefficiencies that leave observable system-level signatures, instead of reporting dormant inefficiencies that have no measurable impact on the current workload.

\section{Discussion}\label{sec:discussion}

\noindent\textbf{Operational implications.}  
A practical challenge in large-scale cloud deployment is that operators can access hardware but not guest workloads.
\unveil demonstrates that meaningful attribution remains possible under this constraint.
Our results show that many silent inefficiencies leave identifiable signatures through recurring contributor patterns from standard system telemetry.
This allows \unveil to localize and correct performance degradations, while maintaining monitoring-grade overhead.

\noindent\textbf{Generality and extensibility.}
\unveil generalizes across both diverse computing architectures and hardware generations.
Although our evaluation primarily targets x86-64 CPUs and NVIDIA GPUs, the selected system-level metrics have equivalent counterparts across other platforms, including Intel, AMD, ARM, and RISC-V.
Since \unveil works with raw telemetry without architecture-specific assumptions, it is portable across hardware.
Despite the generational gap between NVIDIA Volta and ADA, \unveil captured anomaly patterns effectively without extra tuning.
Although anomalies and bottlenecks manifest differently across hardware generations and accelerators, e.g., SM utilization or memory bandwidth saturation on GPUs, these signals remain semantically equivalent at the subsystem level.
For architectures such as ARM CPUs with specific performance monitoring unit (PMU) counters, it can be easily incorporated using the telemetry register (\autoref{sec:implementation}).

\noindent\textbf{Online detection and attribution.}
\unveil performs online anomaly detection using sliding windows and an ensemble of unsupervised models.
Our novel fingerprint technique further localizes known inefficiencies from a knowledge base.
Although the knowledge base needs to be generated, its fingerprints do not rely on a quantitative baseline but capture an inefficiency's qualitative characteristics.
The current metric set used in \unveil is relatively balanced across subsystems.
However, future new hardware and workloads may require additional metrics for some subsystems.
This results in uneven numbers of metrics between them, which may cause bias in some detectors.
For example, subsystems with more metrics may have a larger impact on Mahalanobis distance and Isolation Forest.
However, this is not fundamental, as it can be resolved, e.g., by normalizing their dimensionality with principal component analysis with \unveil metric analysis engine.
We can also expand the fingerprint algorithm to enforce coverage between subsystems, rather than just \emph{metric classes}.

\noindent\textbf{Scalability and production deployment.}\label{sec:scalability}
\unveil is extremely lightweight (\autoref{sec:overhead}), and most of the overhead comes from metric sampling.
Despite our metric analysis engine successfully reducing the required metrics by 40\%, it is still affected by the number of host-visible metrics.
An increase in hardware parallelism (e.g., the number of CPU cores can reach hundreds) and device counts (GPUs per node) causes an increase in the number of \emph{metrics}.
Across hosts, the overhead remains the same because \unveil is executed independently per node, only collaborating when generating reports or cross-validating time series for extra anomaly detection.

\noindent\textbf{Knowledge base evolution.}
We introduce the concept of a knowledge base (\autoref{sec:knowledge-base}), which is a simple key-value store consisting of the fingerprints of known inefficiencies and recipes for remediation.
If a detected anomaly cannot be matched to the knowledge base, and the inefficiency is later confirmed, it is left to the engineer for future analysis.
If an inefficiency is later confirmed and remediation is possible, it can be added to the knowledge base and distributed to benefit future deployments.

\vspace{-5pt}
\section{Related Work}

\noindent\textbf{Performance inefficiencies in ML systems.}
Many studies examine performance inefficiencies in large-scale ML workloads by focusing on specific subsystems or execution phases.
Prior work shows that communication accounts for much of training time, motivating optimizations in collective communication and workload placement~\cite{Huang2024distmm,luo2024disaggregated,rajasekaran2024cassini}.
Other studies characterize storage burstiness~\cite{zoudissecting}, GPU underutilization caused by I/O stalls~\cite{hu2024characterizationlargelanguagemodel}, memory-access imbalance~\cite{legtchenko2025managed}, and performance variability in large accelerator deployments~\cite{sinha2022not}.
These works primarily characterize particular bottlenecks, whereas \unveil automatically detects and attributes silent inefficiencies across hardware subsystems.

\noindent\textbf{Mechanisms behind silent inefficiencies.}
NUMA locality affects GPU communication and RDMA performance~\cite{li2020evaluating,hu2025demystifying,nelson2015numa}, while interrupt affinity governs network throughput~\cite{pesterev2012improving,belay2014ix}.
However, they are treated as the responsibility of the workload owner rather than the operator.
Huge-page fragmentation increases address-translation overhead and degrades memory performance~\cite{basu2013efficient,kwon2016coordinated,panwar2019hawkeye}.
Recommendation inference is particularly sensitive to it because of embedding-table lookups~\cite{ke2020recnmp}.
\unveil detects this during runtime, after it manifests as measurable inefficiencies on the application.
GPU power and frequency management are also studied extensively for optimization.
An operator chooses a power cap or frequency target to save energy, with full knowledge of its throughput cost~\cite{patel2024power,stojkovic2025dynamoLLM,you2023zeus}.
\unveil treats this as an inefficiency by capturing its effect on application performance, regardless of whether it is an inefficiency or an unexpected effect due to operator policy.

\noindent\textbf{Monitoring and diagnosis systems.}
Section~\ref{sec:backgroundmotivation} and \autoref{tab:related_monitor_comparison} compare existing monitoring and diagnosis systems in detail.
\unveil instead creates its own lightweight telemetry register to ensure portability and low dependency when deploying across systems.
They differ in their deployment assumptions, reference requirements, and attribution granularity.
More specifically, a majority require instrumenting the guest-side environment.
\unveil relies on host-visible metrics for within-run anomaly detection.
After anomaly windows are detected, remediation is identified through a fingerprinting process.
Tools such as Prometheus~\cite{turnbull2018monitoring} and LDMS~\cite{osti_1106397} perform telemetry collection and invoke user-defined actions.

\vspace{-5pt}
\section{Limitations and Future Work}
\unveil currently uses fixed sampling rates and static windows.
Although sufficient for the workloads studied here, adaptive sampling and dynamic window sizing may improve responsiveness to changing workloads.
\unveil focuses on runtime inefficiency, which may miss stable inefficiencies present from the start. Related work  on pre-deployment validation~\cite{superbench24} can help address these.
Our detection accuracy analysis (\autoref{tab:detector_perf}) is limited by a lack of window-level ground truth.
We plan to further investigate model behavior by correlating NVTX annotations, iteration timing, and fault injection time, allowing fine-grained accuracy measurement.
While \unveil can automatically remediate certain runtime-recoverable inefficiencies using knowledge base recipes, some issues still require guest cooperation for workload restart or explicit operator intervention.
However, this is not fundamental, as it can be incorporated into the system's workflow.
When signatures tie, \unveil runs all associated remedies. Although each script is validated individually, their actions may negatively interact with each other.
We leave this as future work, and intend to explore how AI agents can be used to propose recipes, with safeguards for their validation and execution.
The evaluation dataset is available at~\cite{chen2026unveil}.

\vspace{-5pt}
\section{Conclusion}
We presented \unveil, a hardware-centric framework for detecting, attributing, and remediating silent inefficiencies in ML workloads that is suitable for cloud operator-level deployment.
\unveil achieves this using four principles.
Firstly, to rely only on host-visible telemetry to enable operator-level deployment.
Secondly, to perform within-run anomaly detection to ensure portability across workloads and environments.
Thirdly, it matches detections to a curated knowledge base and automatically runs prevalidated remediation scripts, rather than simply isolating problematic nodes.
Finally, its low overhead allows continuous deployment without draining compute or storage resources.
Our evaluation shows that host-visible telemetry is sufficient to detect, attribute, and remediate a broad range of silent inefficiencies while maintaining monitoring-grade overhead.

\begin{acks}
This work was funded by the Advanced Research + Invention Agency (ARIA) through project code NACB-PR01-P04 and supported by the John Fell Oxford University Press Research Fund. We acknowledge the use of University of Oxford Advanced Research Computing (ARC) facility (https://doi.org/10.5281/zenodo.22558) and the Isambard 3 Tier-2 HPC Facility (EP/X039137/1). We thank Gregory Tourte, Tina Friedrich, Steven Young, Yassamine Mather, Tom King, and Ashish Thandavan for IT support.
\end{acks}

\newpage
\bibliographystyle{ACM-Reference-Format}
\bibliography{reference}
\end{document}